%% file: main.tex
\PassOptionsToPackage{numbers,sort&compress}{natbib}
\documentclass[10pt,twocolumn,letterpaper]{article}

 \usepackage{cvpr}              

\input{preamble}

\definecolor{cvprblue}{rgb}{0.21,0.49,0.74}
\usepackage[pagebackref,breaklinks,colorlinks,citecolor=cvprblue]{hyperref}

\def\paperID{*****} 
\def\confName{CVPR}
\def\confYear{2024}

\title{\textit{Purrfessor}: A Fine-tuned LLaVA Diet Health Chatbot}

\author{Linqi Lu, Yifan Deng, Chuan Tian, Sijia Yang, Dhavan Shah\\
University of Wisconsin-Madison\\
{\tt\small [llu84, ydeng89, ctian35, syang84, dshah]@wisc.edu}
}

\usepackage{amsthm}

\usepackage{graphicx}
\usepackage{tabularray}
\usepackage{color}
\usepackage{comment}

\definecolor{myblue}{RGB}{28,73,144}
\definecolor{mygreen}{RGB}{44,85,17}
\definecolor{myred}{RGB}{239,64,64}

\begin{document}
\maketitle
\input{0_abstract}

\input{1_intro}
\input{2_method}

\input{3_discussion}
{
    \small
    \bibliographystyle{ieeenat_fullname}
    \bibliography{main}
}
\vspace{1cm}

\input{x_suppl}

\end{document}

%% file: preamble.tex
\usepackage[dvipsnames]{xcolor}


%% file: 0_abstract.tex
\begin{abstract}
This study introduces \textit{Purrfessor}, an innovative AI chatbot designed to provide personalized dietary guidance through interactive, multimodal engagement. Leveraging the Large Language-and-Vision Assistant (LLaVA) model fine-tuned with food and nutrition data and a human-in-the-loop approach, \textit{Purrfessor} integrates visual meal analysis with contextual advice to enhance user experience and engagement. Two studies were conducted to evaluate the chatbot’s performance and user experience: (a) simulation assessments and human validation examined the performance of the fine-tuned model; (b) a 2 (Profile: Bot vs. Pet) by 3 (Model: GPT-4 vs. LLaVA vs. Fine-tuned LLaVA) experiment revealed that \textit{Purrfessor} (\textit{Pet + Fine-tuned LLaVA}) significantly enhanced users’ perceptions of care ($\beta = 1.59$, $p = 0.04$) and interest ($\beta = 2.26$, $p = 0.005$) compared to a GPT-4 bot. Additionally, user interviews highlighted the importance of interaction design details, emphasizing the need for responsiveness, personalization, and guidance to improve user engagement.
\end{abstract}

%% file: 1_intro.tex
\section{Introduction}
\label{sec:intro}

Chatbots are increasingly shaping healthcare and lifestyle management solutions in the evolving field of artificial intelligence (AI)-powered conversational agents. The progression of chatbot technology, especially in health management, has transitioned from simple, rule-based systems to sophisticated conversational agents powered by natural language processing (NLP), machine learning (ML), and large language models (LLMs). This shift has enabled chatbots to engage in real-time, context-sensitive interactions, offering advice tailored to individual behaviors. While traditional apps (e.g., MyFitnessPal) provide static tracking of health metrics, AI-enhanced chatbots have the potential to actively guide users through interactive and personalized experiences that adapt over time \cite{laranjo2018conversational, klopfenstein2017bots}. AI-chatbots designed for health interventions offer immense potential, especially in addressing dietary habits and promoting physical activity, two key behaviors linked to chronic conditions such as cardiovascular disease, type 2 diabetes, and obesity \cite{farhud2015lifestyle, cecchini2010health}.

Despite the proliferation of general health tracking apps, many, such as MyFitnessPal and Lifesum, fall short in offering interactive and personalized health dialogues that adapt in real-time to user behaviors or contexts. Recent studies emphasize the potential of AI-driven health chatbots in providing on-demand guidance, including multimodal inputs, overcoming limitations associated with static health apps and accessibility barriers \cite{pew2017voiceassistants, zhang2020chatbotdiet}. For example, leveraging computer vision alongside NLP, chatbots can now assess user-uploaded images of meals to provide immediate feedback, a capability that elevates their role in dietary management \cite{maher2020healthcoach, salvador2017cooking}.

Building on this work, this study introduces an innovative AI chatbot, \textit{Purrfessor}, designed with advanced interactive functionalities, including personalized feedback facilitated by the LLAVA model (LLaVA-v1.6-13b) \cite{liu2024instruction} fine-tuned through instructional datasets, to provide personalized dietary guidance based on user-uploaded food images and text prompts. With a structured data approach that maps food types to nutritional information and recipe suggestions, the chatbot offers users customized meal analysis to enhance the relevance and accuracy of nutritional assessments \cite{nazary2023healthprompt, salvador2017cooking}. Our exploration of this chatbot's interactive design aims to reveal insights into how AI chatbots can become compelling health companions, fostering positive dietary habits and potentially reshaping the landscape of AI health interventions \cite{maher2020healthcoach, bickmore2005relational}.

One of the primary goals of the \textit{Purrfessor} system is to address pressing dietary and health-related challenges faced by specific populations, particularly those in low socioeconomic status (SES) groups and historically disadvantaged groups where time and resource constraints can hinder healthy meal planning. By providing accessible, rapid suggestions for balanced meals based on ingredients users have on hand, \textit{Purrfessor} is positioned to support families with limited grocery budgets or time to prepare nutritious meals, offering a tailored solution to concerns like childhood obesity and nutrition insufficiencies common in many low-SES and/or historically disadvantaged communities \cite{han2023chatbots, fadhil2017chatbot}.

This research further examines user experiences of this multimodal chatbot, focusing on engagement, emotional attachment, and the perceived efficacy of its meal analysis capabilities. By simulating relational dynamics within chatbot interactions, the system seeks to foster emotional engagement and satisfaction, which are critical for sustained behavioral change of dietary habits \cite{bickmore2005relational, pereira2019healthchatbots}. This study seeks to inform the design of AI chatbots that go beyond static advice, providing a blueprint for a conversational health assistant capable of supporting ongoing lifestyle modifications in response to dynamic individual contexts and resources \cite{zhang2020chatbotdiet, miner2020covid}.

%% file: 2_method.tex
\section{System Design and Structure}
\label{sec:data}
\subsection{System Architecture for Testing}
\begin{figure}[h]
\centering
\includegraphics[width=\linewidth]{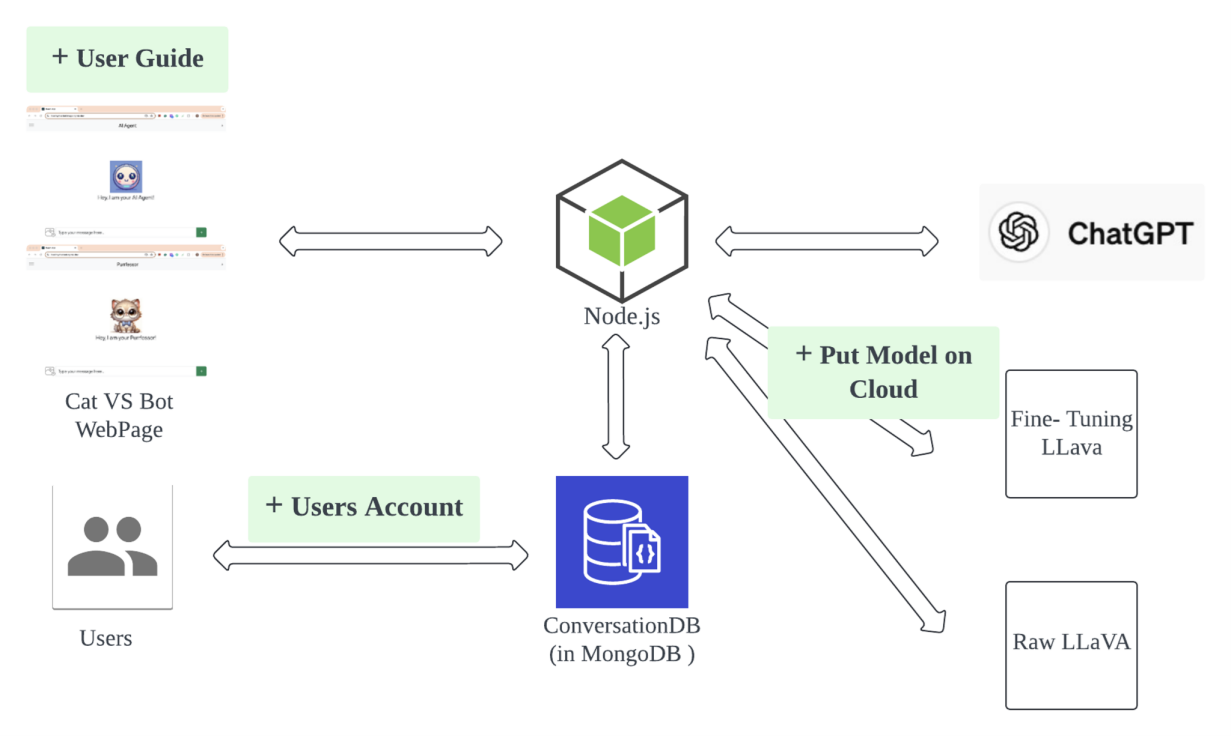} 
\caption{System Architecture.}
\label{fig:system_architecture}
\end{figure}

The system architecture for the AI-powered dietary chatbot, illustrated in Figure~\ref{fig:system_architecture}, is designed to facilitate seamless interaction between users and an advanced conversational AI model, with a focus on dietary guidance and health recommendations. The architecture leverages cloud-based resources, a structured database, and front-end web applications to create an interactive, user-friendly experience.

\subsection{Components}
\subsubsection{User Interface}
The main user interface is accessible via a web page, referred to as \textit{“Pet vs Bot”}. This webpage serves as the primary interaction point for users, allowing them to converse with the chatbot, upload images for analysis, and receive dietary advice.

\subsubsection{User Accounts Management}
The system includes a user account feature that enables personalized interactions and user-specific recommendations. User account data is stored and managed within a secure environment, allowing the chatbot to remember user preferences, past interactions, and provide tailored guidance.

\subsubsection{Server}
Node.js serves as the central hub of the system, orchestrating the flow of data between the front end, database, and AI models. The Node.js server is responsible for:
\begin{itemize}
    \item Handling HTTP requests from the user interface and sending responses back to the webpage.
    \item Processing API requests to interact with the ChatGPT and fine-tuned LLaVA models, facilitating real-time responses.
    \item Routing data to and from the \textit{ConversationDB}, ensuring smooth storage and retrieval of user conversations.
\end{itemize}

\subsubsection{Conversation Database (ConversationDB in MongoDB)}
MongoDB is used as the database management system to store user interactions and chat history. Known as \textit{ConversationDB}, this database enables the system to:
\begin{itemize}
    \item Log user inputs and chatbot responses to ensure continuity in user interactions by referencing past conversations.
    \item Analyze user behavior to refine chatbot responses and enhance personalization.
    \item Support model fine-tuning by storing real user interaction data for improving chatbot accuracy over time.
\end{itemize}

\subsubsection{Cloud-Based Model Hosting}
The fine-tuned LLaVA model is deployed on a cloud server, allowing the system to perform complex computations and ensure scalability. The cloud-based hosting provides:
\begin{itemize}
    \item Efficient processing of image data and response generation for multiple concurrent users.
    \item Scalability to handle increased demand without compromising performance.
    \item Model updates and improvements, enabling the team to fine-tune and enhance response accuracy over time.
\end{itemize}

\subsection{Chatbot Profile}
The \textit{Purrfessor} chatbot’s visual persona is represented by a cat-themed avatar with large, inviting eyes, glasses, and a bowtie, projecting a scholarly yet charming personality. This profile is designed to create a warm and engaging experience, encouraging users to return for guidance on meal planning.

\subsection{Interface Design and Interaction Flow}
\subsubsection{Conversation Bar with Prompt Suggestions}
The interface displays a set of prompt suggestions just above the conversation bar to inspire user interaction, especially for new users. Examples include:
\begin{itemize}
    \item \textit{“Please take a look at my refrigerator and tell me what healthy meals I can cook!”}
    \item \textit{“Have a dish in mind? Upload an image, and we’ll find a matching recipe.”}
    \item \textit{“Looking for a Seafood Pasta recipe today?”}
    \item \textit{“Want to explore high-protein vegetarian meals?”}
\end{itemize}
These prompts can be clicked to auto-populate the conversation bar, streamlining the user experience.

\subsubsection{Menu Icon Hints with Tooltip-Based Approach}
To improve navigation, the menu includes tooltip hints that appear when users hover over or tap on icons. For example:
\begin{itemize}
    \item Hovering over the \textit{“Meal Plan”} icon reveals the tooltip: \textit{“Click here to generate a custom meal plan based on your preferences.”}
    \item Other features include a “?” icon, allowing users to view a short description of their purpose.
\end{itemize}

\subsubsection{Cue for Uploading Images}
An interactive image upload button is integrated into the conversation bar for features involving image-based recipe suggestions. This button changes color or provides visual cues to encourage users to upload images of ingredients or meals.

\subsubsection{Interactive Walkthrough on First Use}
New users are guided through the main features of the chatbot via an interactive walkthrough. This onboarding process highlights:
\begin{itemize}
    \item Key interface elements, such as prompt suggestions, the image upload feature, and menu icons.
    \item Demonstrations on how to start a conversation by selecting example prompts or typing custom questions.
\end{itemize}

\subsubsection{Menu Functions}
\begin{itemize}
    \item \textbf{Chat Interface:} Utilizing image recognition capabilities, \textit{Purrfessor} identifies foods in user-provided images, assessing nutritional indices to offer personalized food recommendations.
    \item \textbf{Misinformation Clarification:} By integrating verified information from authoritative sources, \textit{Purrfessor} addresses common misconceptions related to diet and nutrition, promoting accurate knowledge.
    \item \textbf{Recipe Showcase:} \textit{Purrfessor} displays cooking inspiration aligned with dietary preferences and nutritional needs.
\end{itemize}

\begin{figure}[h]
\centering
\includegraphics[width=\linewidth]{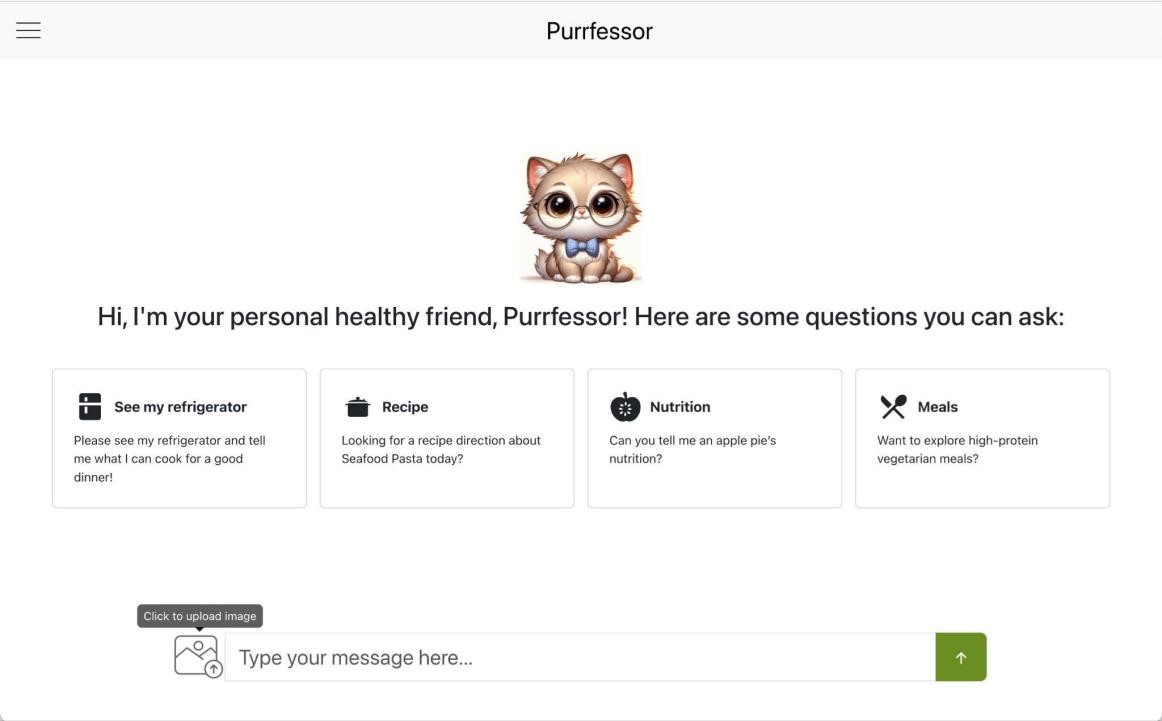}
\caption{\textit{Purrfessor} Nov 24 Version User Interface}
\label{fig:interface_new}
\end{figure}

\section{Chatbot Fine-tuning Methods}
LLaVA is a refined large language model (LLM) that integrates \textit{“an open-set visual encoder from CLIP with the Vicuna language decoder”} to bridge visual and linguistic understanding in multimodal contexts \cite{liu2024instruction}. This architecture allows LLaVA to process and interpret both images and text, enabling it to handle diverse instruction types—such as image descriptions, dialogue, and complex reasoning—by leveraging the strengths of both its visual and language model components. During our experiments, we used the LLaVA-Vicuna-13b 1.6 version as the starting point. Since LLaVA has not been specifically trained on food and nutrition-related data, we performed fine-tuning using a custom training dataset. 

\subsection{Training Dataset}
The training dataset was constructed by integrating multiple data sources:
\begin{enumerate}
    \item \textbf{FoodData Central: Foundation Foods} from the U.S. Department of Agriculture (USDA) (\url{https://fdc.nal.usda.gov/}).
    \item \textbf{Recipe1M dataset} \cite{salvador2017cooking}: This dataset, designed for learning cross-modal embeddings between cooking recipes and food images, includes cooking instructions, nutrition data, and ingredients. It was used to construct the instruction tuning dataset.
    \item \textbf{Human-Annotated Dataset}: A dataset of 500 human-annotated food-related images was included to enhance the model's capability to provide healthy cooking instructions starting from visual inputs of ingredients.
\end{enumerate}

To ensure the model could generate precise nutritional information, we incorporated the \textit{FoodData Central: Foundation Foods} dataset and conducted knowledge injection into the model. The instruction tuning dataset also included caring and supportive language at the beginning and end of responses to enhance the emotional support provided to users.

\subsection{Human-annotated Data}
\subsubsection{Image Collection Using Google Image Search}
A Google Image Search (GIS) method was implemented to systematically collect food-related images from online sources using Python and the Google Custom Search API. This approach follows established practices in data-intensive research for image retrieval \cite{zhu2021imagesearch, redmon2018yolov3}. The collection process utilized the \texttt{googleapiclient} package \cite{google2023customsearch}, with a custom search engine ID and API key configured to automate retrieval. 

A set of pre-defined search queries, tailored with keywords related to raw produce and cooking ingredients, was employed to filter relevant images and metadata. Example queries included:
\begin{itemize}
    \item \textit{“raw food”}
    \item \textit{“produce and meat in fridge”}
    \item \textit{“fresh produce”}
    \item \textit{“produce in fridge”}
    \item \textit{“food ingredients”}
    \item \textit{“raw meat and produce”}
    \item \textit{“cooking raw meat and vegetables”}
    \item \textit{“fresh produce for cooking”}
\end{itemize}

Metadata extracted for each image included:
\begin{itemize}
    \item Collection date
    \item Page title
    \item Source website
    \item Image URL
    \item Page URL
\end{itemize}
Where available, dates were extracted directly from URLs using regular expressions, following established methodologies for metadata organization \cite{beck2020webscraping}. This enabled chronological organization of the dataset for subsequent analysis and validation.

\subsubsection{Image Captioning and Q\&A Generation}
Each image in the dataset was processed through GPT-4o using a structured prompt designed to elicit detailed captions and chatbot-friendly Q\&A examples. The prompt instructed the model to generate the following outputs:
\begin{itemize}
    \item A detailed caption describing visible ingredients.
    \item A Q\&A response including:
    \begin{itemize}
        \item A greeting.
        \item Nutritional information for the identified ingredients.
        \item Healthy recipe suggestions.
        \item A step-by-step guide for each recipe.
        \item A closing message encouraging user engagement.
    \end{itemize}
\end{itemize}

To ensure consistency, each prompt followed a structured output format, facilitating straightforward parsing and subsequent training of the LLaVA model. The inclusion of nutritional information and recipe suggestions aimed to simulate real-world interactions, enabling the model to respond both informatively and engagingly \cite{radford2021clip}.

\subsubsection{Human-in-the-Loop}
After the initial output generation, human annotators reviewed each caption and Q\&A example to ensure clarity, accuracy, and alignment with expected chatbot responses. Key aspects of the review process included:
\begin{itemize}
    \item Verifying language appropriateness and grammatical correctness.
    \item Ensuring logical flow and alignment with user expectations.
    \item Handling edge cases and special characters appropriately.
\end{itemize}

For ambiguous or non-food images, annotators crafted captions and Q\&A responses neutrally, avoiding refusals and maintaining factual descriptions of the content in line with the prompt guidelines \cite{zhu2021imagesearch}. Edge cases were explicitly included in the dataset, with Q\&A examples refined by human reviewers to ensure robust training data. This approach prepared the model to handle diverse real-world scenarios effectively.

\subsection{Fine-Tuning Strategy}
Initially, we attempted to perform full tuning on the LLaVA-Vicuna-13b model, as the abundance of examples in the Recipe1M dataset provided a strong foundation. However, the GPU memory required to fully tune a 13B-parameter model exceeded the capacity of the available hardware. To address this limitation, we adopted the Low-Rank Adaptation (LoRA) technique, which enables fine-tuning of large models with reduced computational requirements.

While LoRA proved effective for reducing the hardware burden, we encountered overfitting issues with the Recipe1M dataset. Specifically, the model tended to respond with cooking instructions for every query, regardless of context. To mitigate this, we diversified the training set by integrating the human-annotated visual dataset, enabling the model to provide healthy cooking instructions starting from visual inputs, such as raw ingredients.

The fine-tuned version of the \textit{Purrfessor} chatbot represents a state-of-the-art approach in AI-driven dietary guidance. By combining advanced multimodal technology with user-centric design principles, the chatbot offers evidence-based nutritional information, personalized recipe suggestions, and emotionally supportive interactions, setting a new benchmark for dietary chatbots.

\section{Study A. Chatbot Simulation Testing}
\subsection{Methodology}
To evaluate the performance of our fine-tuned chatbot, \textit{Purrfessor}, we adopted a two-phase evaluation framework that combined automated metrics and human-in-the-loop validation. This approach ensured a rigorous assessment of both quantitative and qualitative aspects of the model's performance.

\subsubsection{Dataset and Q\&A Generation}
The dataset for this study consisted of 500 real-world images sourced via the Google Image Search API, paired with synthetically generated prompts designed to simulate a wide range of user queries. The images encompassed diverse food categories, including everyday meals, beverages, and culturally specific dishes, reflecting the variety users might encounter in real-world applications. Prompt diversity included direct identification tasks (e.g., \textit{“What food-related items appeared in this image?”}) and contextual queries (e.g., \textit{“Do you have ideas on recipes with low-fat per the current ingredients I provide?”}). This ensured that the Q\&A pairs captured both fundamental object detection and contextual reasoning challenges.

\subsubsection{Evaluation Metrics}
For automated assessment, we employed a text overlap metric to measure consistency between detected food-related nouns in chatbot responses and expected terms. Using an NLP-based approach, we extracted food-related words and analyzed the overlap rate among food-related nouns from GPT-4 and \textit{Purrfessor}. This overlap score served as an initial indicator of accuracy, with a threshold of 0.6 used to identify low-performing cases. Recognizing the limitations of simple overlap-based metrics in capturing semantic nuances (e.g., \textit{"greens"} vs. \textit{"lettuce"}), we adopted a human-in-the-loop approach to address these gaps. Future studies could incorporate semantic similarity measures, such as cosine similarity of word embeddings, to provide a more robust evaluation.

\subsubsection{Human Validation Process}
Human validation was conducted by three human coders who assessed the responses from the fine-tuned LLaVA model, \textit{Purrfessor}, using four main criteria (see Table 1): 
\begin{enumerate}
    \item \textbf{Correctness}: Accuracy of food item detection in the images.
    \item \textbf{Relevance}: Pertinence of the chatbot's responses to the user queries.
    \item \textbf{Clarity}: Coherence and ease of understanding of the responses.
    \item \textbf{Handling Edge Cases}: Ability to address ambiguous or non-food images appropriately.
\end{enumerate}

Each criterion was rated on a 10-point scale following established methodologies for scaled evaluation in chatbot performance studies \cite{pereira2019healthchatbots}. Detailed scoring guidelines were provided to ensure consistency (see Appendix I). For example, \textbf{correctness} was evaluated based on the accurate identification of food items in an image, while \textbf{handling edge cases} examined responses to ambiguous or non-food images. 

Inter-coder reliability was calculated using Krippendorff’s alpha, yielding high agreement among evaluators:
\begin{itemize}
    \item $\alpha_{\text{correctness}} = 0.86$
    \item $\alpha_{\text{relevance}} = 0.96$
    \item $\alpha_{\text{clarity}} = 0.91$
    \item $\alpha_{\text{edge case}} = 0.85$
\end{itemize}

The responses were categorized using a three-level scale: \textit{low} (1–3), \textit{medium} (4–7), and \textit{high} (8–10). These scores provided a comprehensive evaluation of the chatbot's performance, highlighting its strengths and areas for improvement.

\subsection{Results}
\subsubsection{Automated Metrics}
The text overlap score for image object detection tasks averaged 0.67, indicating moderate alignment between the chatbot’s outputs and expected responses. This score reflects performance differences between GPT-4 and the fine-tuned LLaVA model. Discrepancies in low-scoring cases often arose from nuanced distinctions (e.g., \textit{“pasta”} vs. \textit{“spaghetti”}) or the omission of less visually salient items (e.g., garnishes). To address these limitations, a human review of low-overlap cases (\textless 0.6, $n=100$) was conducted to validate the results and provide qualitative insights into performance gaps.

\subsubsection{Comparison with GPT-4}
A comparison between \textit{Purrfessor} and GPT-4 highlighted differences in descriptive depth and contextual richness. While GPT-4 often provided elaborative responses that included nutritional context (e.g., \textit{“a vibrant dish high in fiber, featuring fresh vegetables and whole grains”}), \textit{Purrfessor} prioritized concise and factual outputs (e.g., \textit{“vegetable salad with carrots and spinach”}). Although the succinct responses from \textit{Purrfessor} were generally accurate, they provided fewer contextual details that users might find helpful in decision-making scenarios.

\subsubsection{Human Evaluation}
The human validation scores demonstrated the chatbot’s strengths and areas for improvement across the four evaluation criteria:

\paragraph{Correctness ($M=7.87$):}
Responses were largely accurate, though minor discrepancies were noted in images containing overlapping or visually similar items (e.g., distinguishing \textit{“arugula”} from \textit{“green lettuce”}). Obscure objects were often not specified in the responses.

\paragraph{Relevance ($M=9.4$):}
Chatbot responses effectively addressed user queries, though context-specific prompts occasionally received overly generic answers.

\paragraph{Clarity ($M=9.6$):}
Responses were clear, structured, and reflected the training prompt format. However, occasional incomplete outputs were noted due to token limitations.

\paragraph{Handling Edge Cases ($M=9.0$):}
The model effectively recognized ambiguous inputs, such as non-food images, often providing disclaimers like \textit{“This image does not appear to contain food.”} Edge cases were categorized into two main types:
\begin{enumerate}
    \item \textbf{Ambiguous Food Features:} Images with overlapping or partially obscured food items, such as a pizza slice buried under toppings. While \textit{Purrfessor} accurately identified most major ingredients, low overlap scores often resulted from the omission of secondary items (e.g., basil leaves).
    \item \textbf{Non-Food Images:} The chatbot successfully flagged non-food-related images with appropriate disclaimers, demonstrating its robustness in unexpected scenarios.
\end{enumerate}

Future iterations of the model could further improve performance by incorporating a secondary verification step for ambiguous classifications.

\section{Study B. User Experience Testing}

\subsection{Study Design}
This study utilized a 2 (Chatbot Profile: Bot vs. Anthropomorphic) by 3 (Chatbot Model: GPT-4 vs. Raw LLaVA vs. Fine-tuned LLaVA) between-subject design, with an additional Baseline condition (ChatGPT only). The independent variables were visual profile (Bot, Anthropomorphic) and chatbot algorithm (ChatGPT-4, Fine-tuned LLaVA), while the dependent variables included user experience, attitudes, and intentions to adhere to the chatbot-recommended recipes. Participants interacted with the chatbot by asking questions, uploading photos, and receiving personalized healthy recipe recommendations.

\subsection{Data Collection and Participants}
Participants were undergraduate students aged 18-25 (\emph{N} = 51) from a Midwestern university who received extra credit for their participation. Each participant was randomly assigned to one of six experimental conditions. Prior to the experiment, participants provided informed consent and completed a pre-experiment questionnaire collecting demographic information, assessing technology literacy, and gauging their predisposition toward AI-powered chatbots. 

Participants engaged in a 15-minute conversation with their assigned chatbot, during which they received personalized dietary recommendations and guidance. Post-interaction, they completed a questionnaire evaluating user experience, attitudes toward the chatbot, engagement levels, and perceptions of the sequenced conversation. At the study's conclusion, participants were debriefed about the chatbot’s information sources and thanked for their participation.

\subsection{Measures}
\paragraph{Manipulation Check.}
Two questions were designed to verify participants' recall of the chatbot they interacted with during the experiment. Participants were asked to identify the chatbot's name and describe its profile image.

\paragraph{User Experience Questionnaire (UEQ).}
The User Experience Questionnaire (UEQ) evaluated user interactions with the chatbot, covering aesthetic impressions and functional effectiveness \cite{schrepp2023importance}. Participants rated eight items (\emph{M} = 3.69, \emph{SD} = 0.71, $\alpha$ = .92). Key dimensions included:

\begin{itemize}
    \item \textbf{Attractiveness:} Assessed overall aesthetic appeal (e.g., “annoying/enjoyable,” “attractive/unattractive”).
    \item \textbf{Perspicuity:} Evaluated ease of use (e.g., “difficult to use,” “clear/confusing”).
    \item \textbf{Stimulation:} Determined the product's excitement and motivation to use (e.g., “boring/exciting”).
    \item \textbf{Novelty:} Gauged innovation and creativity (e.g., “conservative/innovative”).
\end{itemize}

\paragraph{Care Perceptions.}
Care perceptions were measured using four items rated on a 5-point Likert scale (\emph{M} = 4.97, \emph{SD} = 1.33, $\alpha$ = .91). Example items included: “This chatbot cares about our well-being” and “This chatbot makes me feel a kind of emotional support.”

\paragraph{User Interest.}
User interest was assessed through three 5-point Likert scale questions (\emph{M} = 4.97, \emph{SD} = 1.33, $\alpha$ = .91). Sample items included: “This chatbot is appealing” and “This chatbot piques my interest in engaging in conversations.”

\paragraph{Compliance Intention.}
Compliance intention measured participants' likelihood of adopting chatbot recommendations using three items (\emph{M} = 3.43, \emph{SD} = 0.76, $\alpha$ = .89). Statements included: “I will consider adopting the suggestions from the chatbot that was just used.”

\paragraph{User Satisfaction.}
Satisfaction was evaluated with three statements rated on a 5-point Likert scale (\emph{M} = 3.55, \emph{SD} = 0.85, $\alpha$ = .92). Example statements were: “I am very satisfied with the chatbot system.”

\paragraph{AI Predispositions.}
AI predispositions were assessed for control purposes using two scales:
\begin{itemize}
    \item \textbf{AI Efficacy:} Measured confidence in using AI technologies (\emph{M} = 3.59, \emph{SD} = 0.94, $\alpha$ = .93) \cite{hong2022selfefficacy}.
    \item \textbf{AI Anxiety:} Assessed attitudes toward learning new technologies (\emph{M} = 2.83, \emph{SD} = 0.85, $\alpha$ = .79) \cite{wang2022aianxietyscale}.
\end{itemize}

\paragraph{Demographics.}
Demographic data, including gender, race, household income, ideology, and education, were collected as covariates.

\subsection{Regression Analysis Results}
The multiple linear regression analysis, controlling for demographic characteristics and predispositions, showed the following:

\begin{itemize}
    \item The fine-tuned LLaVA anthropomorphic chatbot \textbf{Purrfessor} ($\beta = 1.59$, $p = 0.04$) and the raw LLaVA anthropomorphic chatbot ($\beta = 1.58$, $p = 0.02$) were both positively associated with \textbf{care}, showing more effective engagement compared to the GPT-4 bot.
    \item The fine-tuned LLaVA anthropomorphic chatbot \textbf{Purrfessor} ($\beta = 2.26$, $p = 0.01$) and the raw LLaVA anthropomorphic chatbot ($\beta = 2.50$, $p < 0.001$) were positively associated with \textbf{user interest} compared to the GPT-4 bot.
    \item For \textbf{user experience quality}, the fine-tuned LLaVA bot-like chatbot ($\beta = 1.10$, $p = 0.02$) and the raw LLaVA cat chatbot ($\beta = 0.88$, $p = 0.02$) showed slight improvements compared to the GPT-4 bot.
    \item Regarding \textbf{overall satisfaction}, the fine-tuned LLaVA bot-like chatbot emerged as a significant enhancer of satisfaction ($\beta = 1.01$, $p = 0.03$) compared to the GPT-4 bot.
    \item \textbf{Compliance intentions} to the chatbot’s suggestions did not show statistical significance, $F(14, 36) = 1.25$, $p = 0.29$.
\end{itemize}

\begin{figure}[h]
  \includegraphics[width=\linewidth]{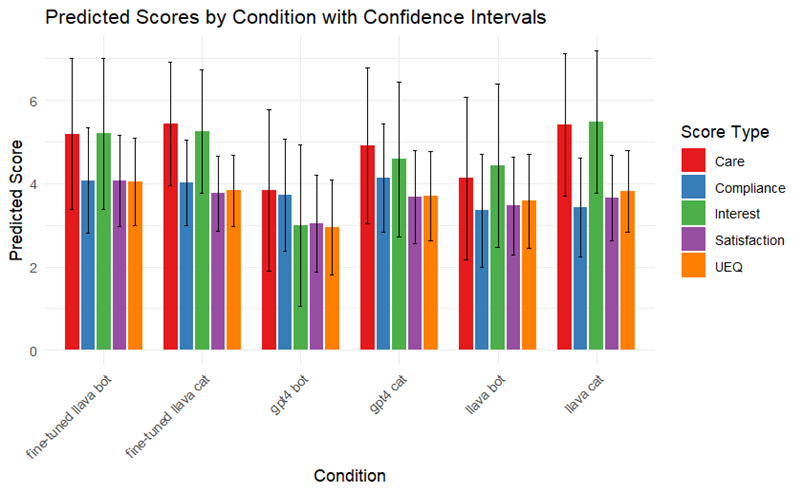}
  \caption{The Multiple Linear Regression Results}
  \label{fig:ue_regression}
\end{figure}

\subsection{User Interview}
An in-depth interview was conducted with 8 participants who interacted with the fine-tuned LLaVA chatbot \textbf{Purrfessor} (in April 24 version). The goal was to gather personal feedback and refine system design. Two interview coders independently reviewed transcripts to ensure consistency in thematic coding, confirmed through cross-validation to enhance accuracy and credibility \cite{braun2006}. A qualitative thematic analysis revealed three primary themes for future improvement: \textbf{Interaction and Responsiveness}, \textbf{Personalization and Relevance}, and \textbf{Guidance and Instructions}.

\subsubsection{Interaction and Responsiveness}
Participants expressed a desire for more seamless and dynamic interactions. Long response times disrupted engagement, reducing the fluidity of conversations. Suggestions included displaying user input immediately and generating chatbot responses progressively to enhance real-time interaction. 

\begin{quote}
    “The waiting time is long. You can make it same time typing to add interaction.”
\end{quote}

\begin{quote}
    “Whenever I ask a question, I have to wait, with my question still in the input box, until the chatbot finishes its response.”
\end{quote}

\begin{quote}
    “The version of output can be improved and the answer format like greetings can add more fun language. Emoji to fit the robot personality.”
\end{quote}

These insights suggest improvements in conversational design, such as real-time responsiveness and playful, context-sensitive language, to enhance user satisfaction and engagement. Adding personality-enhancing language may better align the chatbot’s tone with its anthropomorphic persona, creating a more entertaining and personable experience \cite{braun2006,wiltshire2021}.

\subsubsection{Personalization and Relevance}
Participants emphasized the importance of personalized and context-aware interactions. While chatbot responses were accurate and relevant, the inability to reference previous interactions limited conversational richness.

\begin{quote}
    “The answers seem accurate, useful, and on point; however, when I ask follow-up questions, it does not consider the prior questions I asked.”
\end{quote}

\begin{quote}
    “Adapt the recipe based on the user’s preference and hometown.”
\end{quote}

This feedback highlights the potential value of incorporating memory or contextual awareness features to create interconnected exchanges. Personalization, such as adapting responses based on user preferences or prior interactions, aligns with social interaction theory and can enhance user satisfaction and engagement \cite{braun2019}.

\subsubsection{Guidance and Instructions}
Participants indicated a need for clearer guidance on how to engage with the chatbot effectively. Uncertainty during initial interactions led to hesitation or confusion.

\begin{quote}
    “At the beginning, I didn’t know what to do. If the initial page gave me some hints or introductions, I might be clearer.”
\end{quote}

\begin{quote}
    “You could add some suggestions for users to start a conversation with the chatbot.”
\end{quote}

Introducing conversational prompts or visual guides at the outset may reduce cognitive load and empower users to make full use of the chatbot’s functions. Providing accessible guidance aligns with findings in digital interface studies, which emphasize the importance of clear onboarding processes to facilitate user engagement \cite{sundar2010,wiltshire2021}.

\begin{figure}[h]
  \includegraphics[width=\linewidth]{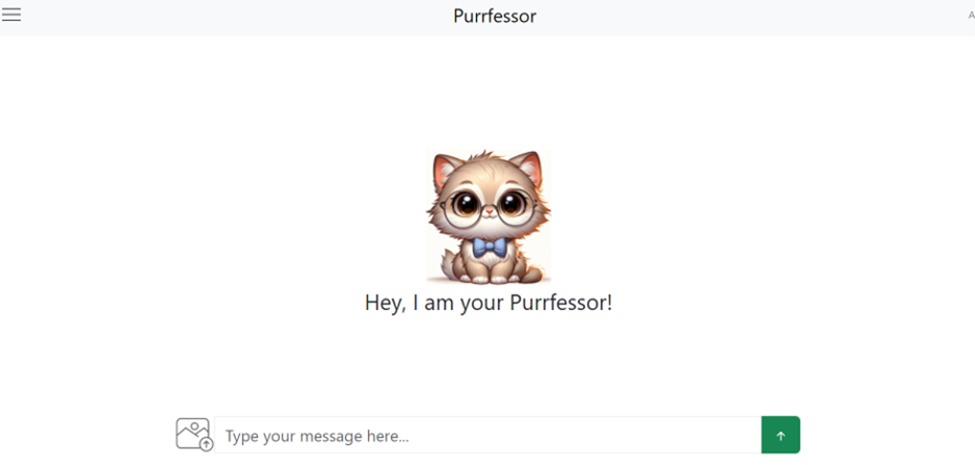}
  \caption{\textit{Purrfessor} April 24 Version User Interface.}
  \label{fig:interface_old}
\end{figure}

%% file: 3_discussion.tex
\section{Discussion and Conclusion}
Our simulation test reveals a distinction in descriptive nuance between GPT-4 and LLaVA. While GPT-4 generated responses that were rich in context, offering nutritional insights and detailed descriptions, LLaVA’s responses were more straightforward and focused on item identification. This difference highlights GPT-4’s strength in generating comprehensive responses, potentially enhancing user understanding and satisfaction. However, LLaVA’s simpler descriptions may appeal to users who prefer direct and accessible information. This discrepancy between the models emphasizes the need to balance narrative richness and simplicity based on user preference and context.

The results suggest that the fine-tuned LLaVA chatbot, designed with an anthropomorphic persona enhanced user perception of care and engagement intentions compared to the GPT-4 chatbot with a bot profile. Both the fine-tuned and raw versions of LLaVA outperformed GPT-4 in eliciting perceptions of care and interest, likely due to the integration of a relatable, visually engaging avatar. This aligns with literature suggesting that personality-driven chatbots can increase user satisfaction and emotional engagement \cite{pereira2019healthchatbots}. However, while Purrfessor performed well in increasing user interest and engagement, the study did not observe significant improvements in compliance behaviors. This suggests that, for health intervention chatbots, simply improving conversational quality or introducing an endearing avatar may not be sufficient to influence behavior adherence in the short term. Future studies could investigate whether sustained interaction with personality-driven bots leads to greater behavioral compliance over time, an area unexplored in current literature.

The qualitative data obtained from user interviews revealed three major themes: interaction responsiveness, personalization, and guidance for effective interaction. Users expressed a desire for quicker response times, suggesting progressive answer displays or real-time typing feedback to enhance the chatbot’s interactivity. This feedback aligns with prior research that emphasizes the role of prompt response in user satisfaction with digital interfaces \cite{han2023chatbots}. Additionally, users sought more personalized interactions, wanting the chatbot to remember prior conversations and adapt to individual preferences, which points toward the importance of adaptive AI systems in fostering long-term user engagement. Finally, users noted that introductory hints or interaction examples would make the chatbot experience more intuitive, underscoring the need for clear guidance to improve first-time user experience.
Despite its insights, this study faces limitations, including limited sample size and the specificity of AI conditions tested, which might not generalize across all types of AI applications. Future research should explore a broader array of AI configurations and include a more diverse demographic to enhance the generalizability of the findings, which could provide deeper insights into the subjective experiences of users interacting with AI systems, complementing the quantitative data presented.

In conclusion, this study advances the field of computational communication by providing concrete evidence on how specific design enhancements—such as personalized visual profiles and refined language-image processing—can significantly impact user engagement and satisfaction in health chatbots. The findings suggest that user-centered design, which incorporates tailored interactions and context-sensitive responses, holds promise for developing chatbots that are not only engaging but also more effective in delivering health-related advice. This study also sets a foundation for future AI experimental research in health chatbot applications, showing that even subtle adjustments in chatbot interface and interaction style can lead to measurable improvements in user experience.

%% file: X_suppl.tex
\clearpage
\setcounter{page}{1}
\maketitlesupplementary

\section*{Appendix A: Validation Criteria}
\label{sec:appendix_a}

\begin{table}[h]
\centering
\caption{Validation Criteria for Chatbot Performance}
\label{tab:validation_criteria}
\renewcommand{\arraystretch}{1.5}
\begin{tabular}{|p{3cm}|p{4cm}|p{6cm}|}
\hline
\textbf{Criteria} & \textbf{Definition} & \textbf{Indicators for Scoring (1-10 scale)} \\ \hline
\textbf{Correctness} & 
Accuracy of image object detection compared to provided text description. & 
\begin{itemize}[noitemsep, topsep=0pt]
    \item \textbf{High (8-10):}Accurate detection of most major food items (80\%-100\%) ; closely matches text description.
    \item \textbf{Medium (4-7):} Partially accurate detection (50\%-79\%); some items missing or misidentified.
    \item \textbf{Low (1-3):} Significant errors(below 50\%); most items inaccurately identified.
\end{itemize} \\ \hline

\textbf{Relevance} & 
Pertinence of the chatbot’s responses to the questions asked. & 
\begin{itemize}[noitemsep, topsep=0pt]
    \item \textbf{High (8-10):} Responses align well with user intent and directly address the question.
    \item \textbf{Medium (4-7):} Address the question but may include extraneous or unrelated details.
    \item \textbf{Low (1-3):} Lack relevance; off-topic or fails to answer the question.
\end{itemize} \\ \hline

\textbf{Clarity} & 
Ease of understanding the chatbot’s responses. & 
\begin{itemize}[noitemsep, topsep=0pt]
    \item \textbf{High (8-10):} Well-organized, grammatically correct, and easy to follow.
    \item \textbf{Medium (4-7):} Generally clear but may contain minor coherence or grammatical issues.
    \item \textbf{Low (1-3):} Poor grammar, lack of organization, or difficult to understand.
\end{itemize} \\ \hline

\textbf{Handling Edge Cases} & 
Effectiveness in managing unexpected or ambiguous queries (e.g., non-food images). & 
\begin{itemize}[noitemsep, topsep=0pt]
    \item \textbf{High (8-10):} Acknowledge ambiguity and provides reasonable responses.
    \item \textbf{Medium (4-7):} Recognize ambiguity but response not clear.
    \item \textbf{Low (1-3):} Fail to handle ambiguous inputs effectively; response is irrelevant or unreasonable.
\end{itemize} \\ \hline

\end{tabular}
\end{table}

\clearpage 

\section*{Appendix B: Data Flow and Operation}
\label{sec:appendix_b}

\subsection*{Data Flow Overview}
The system architecture facilitates seamless interaction between users and the chatbot, ensuring efficient data processing and response generation.

\begin{itemize}[noitemsep, topsep=0pt]
    \item \textbf{User Interaction:} Users interact through the Cat vs Bot webpage, typing questions, selecting prompts, or uploading images.
    \item \textbf{Data Routing:} The Node.js server processes these inputs, routing text and image data to the appropriate model (ChatGPT or LLaVA).
    \item \textbf{Response Generation:} ChatGPT or LLaVA generates responses, which are routed back to the server.
    \item \textbf{Database Logging:} The ConversationDB logs interactions, storing accessible image URLs and using AWS for image storage. This enables personalization and continuity in future interactions.
    \item \textbf{User Response:} Responses are displayed back on the Cat vs Bot webpage, completing the interaction loop.
\end{itemize}

\section*{Appendix C: Conversation Demo}
\label{sec:appendix_c}

\begin{figure}[h]
  \centering
  \includegraphics[width=0.9\linewidth]{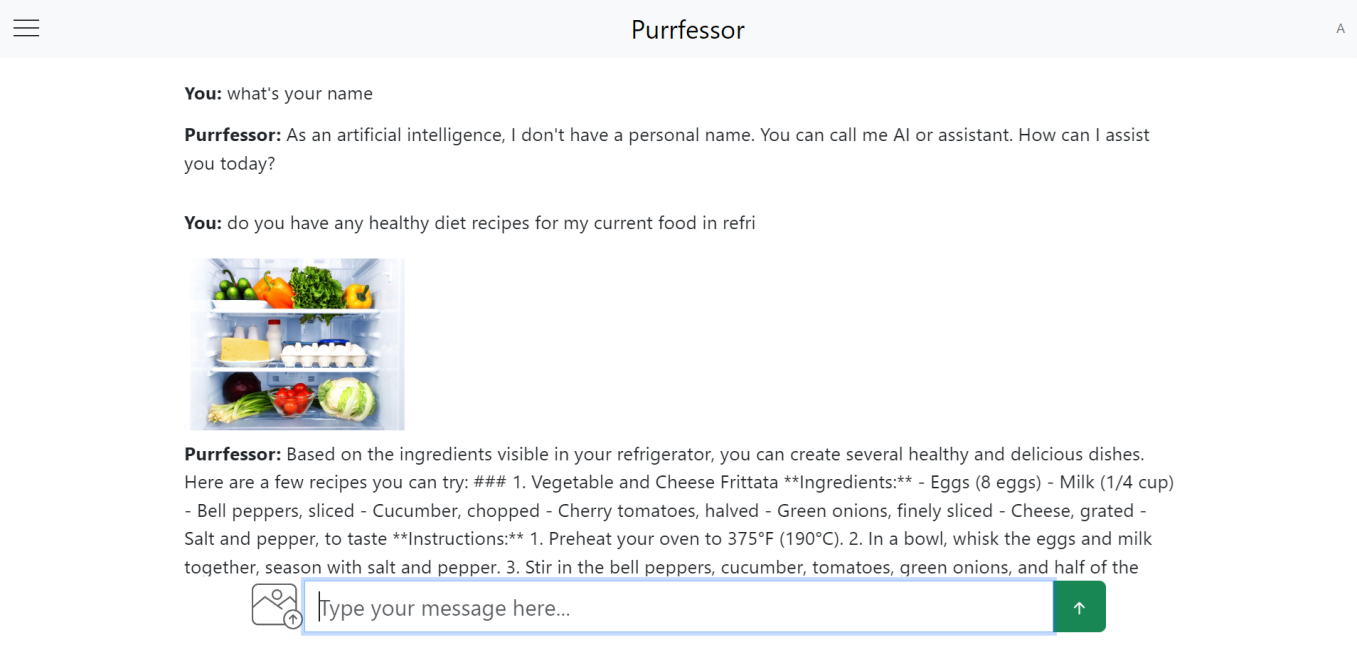}
  \caption{Chatbot Purrfessor Q\&A Conversation Demo.}
  \label{fig:conversation_demo}
\end{figure}